\documentclass{aa}  
\usepackage{graphicx}
\usepackage{longtable}
\usepackage{txfonts}
\begin{document}
   \title{Milliarcsecond angular resolution of
   reddened stellar sources in the vicinity of the 
Galactic Center
\thanks{Based on observations made with ESO telescopes at Paranal Observatory}
}

   \titlerunning{Milliarcsecond angular resolution of
   reddened stellar sources.}

   \author{A. Richichi\inst{1}
          \and
          O. Fors\inst{2}\fnmsep\inst{3}
          \and
          E. Mason\inst{4}
	  \and 
          J. Stegmaier\inst{1}
	  \and 
	  T. Chandrasekhar\inst{5}
          }

   \offprints{A. Richichi}

   \institute{
             European Southern Observatory,
Karl-Schwarzschild-Str. 2, 85748 Garching bei M\"unchen, Germany
             \email{arichich@eso.org}
         \and
             Departament d'Astronomia i Meteorologia, Universitat de  
Barcelona, Mart\'{\i} i Franqu\'es 1, 08028 Barcelona, Spain
         \and
             Observatori Fabra, Cam\'{\i} de l'Observatori s/n, 08035 
Barcelona, Spain
   \and
             European Southern Observatory, Santiago, Chile
   \and
	     Physical Research Laboratory, 380009 Ahmedabad, India
             }


  \abstract
   {}
   {For the first time, the lunar occultation technique has been
   employed on a very large telescope in the near-IR with the aim
   of achieving systematically milliarcsecond resolution on stellar
   sources.
   }
   {
   We have demonstrated the burst mode of the ISAAC instrument,
   using a fast read-out on a small area of the detector to record
   many tens of seconds of data at a time
   on fields of few squared arcseconds.
   We have used the opportunity to record a large number of LO events
   during a passage of the Moon close to the Galactic Center in March 2006.
   We have developed and employed for the first time a
   data pipeline for the treatment of LO data, including the automated
   estimation of
   the main data analysis parameters using a wavelet-based method, 
   and the preliminary fitting and plotting of all light curves.
   }
   {
   We recorded 51 LO events over about four hours. Of these, 30 resulted
   of sufficient quality to enable a detailed fitting.  We detected two
   binaries with subarcsecond projected separation and three stars with a
   marginally resolved angular diameter of about 2 milliarcseconds. Two more
   stars, which are cross-identified with SiO maser, were found to be
   resolved and in one case we could recover the brightness profile of
   the extended emission, which is well consistent with an optically thin
   shell. The remaining unresolved stars were used to characterize the
   performance of the method.
   }
   {
   The LO technique at a very large telescope is 
   a powerful and efficient method to achieve 
   angular resolution, sensitivity, and dynamic range that are among
   the best possible today with any technique. The selection of targets
   is naturally limited and LOs are fixed-time events, however
   each observation requires only a few minutes including overheads.
   As such, LOs are ideally suited to fill small gaps of idle time 
   between standard observations.
   }

   \maketitle
%

\section{Introduction}
The method of observing lunar occultations (LO) of background stars
to derive their angular diameter, as well as other characteristics
such as binarity, has been employed for decades and provides
angular resolution at the milliarcsecond (mas) level 
independently of the diameter of the telescope used.
This surpasses the  diffraction limit
of even the largest single telescopes and rivals the resolution
of long-baseline interferometry (LBI) even with baselines of
hundreds of meters.  Basically, the LO technique 
relies on the lunar limb as a diffracting edge, rather
than on the diameter of the telescope. 
The diffraction fringes are generated in space, and due to their
relatively fast motion
over the telescope they must be sampled at rates of order
0.1-1\,kHz. Combined together, these factors greatly diminish the
adverse effects of atmospheric turbulence, which are the main
limit of other high angular resolution techniques.
LO also permit not only a model-dependent derivation of source
parameters such as the angular diameter and binary parameters, but
also a model-independent reconstruction of the brightness profile of
the source according to maximum-likelihood, or even a unique
reconstruction by light curve inversion in special cases.
The time required for observation (dominated by overheads, since an occultation
lasts much less than 1\,s) and for data analysis is significantly
shorter than for most other high angular resolution methods.

Of course, LO suffer from a number of significant drawbacks, first among them
that the sources cannot be chosen at will. The Moon covers only about 10\%
of the celestial sphere in its apparent orbit. LO are fixed time events,
and as such subject to weather and instrumental downtimes.
Finally, the lunar limb only provides a 1-D scan of the source, although
observations from different sites, or at different epochs, can in
principle be combined under favorable circumstances.
Due mainly to the chromatic properties of the scattered
light background around the Moon and of 
the diffraction fringes, the near-IR is ideally suited
to observe LO.
More details on the method, its performance, the data analysis and the
results can be found in Richichi (\cite{AR_thesis}, \cite{richichi96}) and
Fors (\cite{thesis}). The CHARM2 catalogue alone (Richichi et al.
\cite{charm2}) lists
1815 LO results in the field of high angular resolution.

While the angular resolution of LO is set mainly by the lunar limb
rather than by telescope size, this latter parameter obviously has
a crucial role in determining the limiting sensitivity.
Richichi (\cite{1997IAUS..158...71R})
studied the performance of LO under a number of
circumstances, including the then-futuristic use of IR array detectors
on a 8-m class telescope. This has recently become reality:
Richichi et al. (\cite{Messenger2006}) 
and Fors et al. (\cite{SEA2007}, \cite{fors08})
reported in a preliminary form
on the use of the ISAAC instrument in the so-called burst mode
at the ESO VLT 8.2\,m Antu telescope. 

In the present paper we provide a detailed account of the observations
carried out during a few hours in the night of March 21, 2006,
when 51 LO events were recorded. 
A second batch of observations was carried out in August 2006:
due to their large number and their different nature,
these sources will be discussed in a separate paper.
In Sect.~\ref{data} we describe
the observations and we provide details 
of the data reduction. 
In Sect.~\ref{results} 
we discuss the results, which 
include new binaries, resolved stars and the
near-IR counterparts of two radio maser sources.
We also characterize the performance of this
specialized observing mode, which
combines the highest angular resolution and
sensitivity possible today in the near-IR.
In the conclusions we
describe the integration of the LO technique in the
service mode operations scheme in place at the Paranal observatory
and our plans for routine LO observations at that site.


\section{Observations and data analysis}\label{data}
\begin{figure*}
\includegraphics[width=18cm]{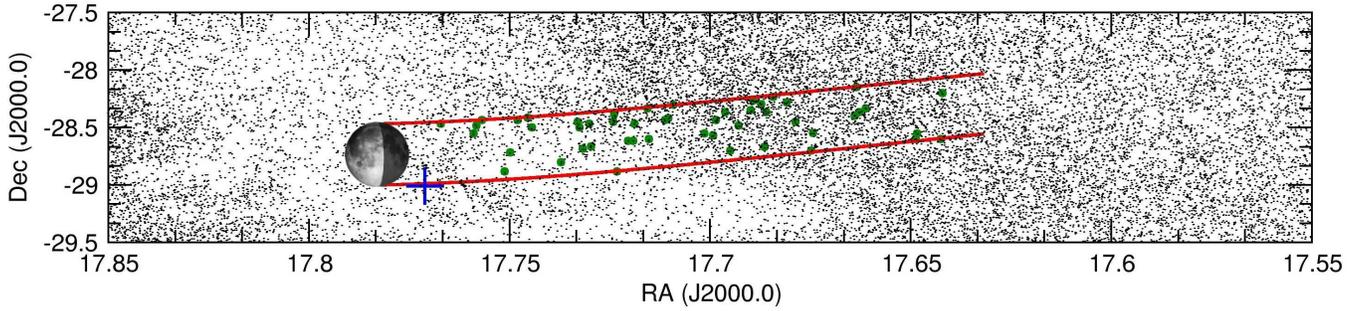}
\caption {The area close to the sources occulted by the
Moon in the night of March 21, 2006. The small dots represent
sources in the 2MASS catalogue with magnitude K$\le 7.5$. The heavier
dots are the sources for which we could record a lunar occultation.
The apparent path of the Moon, moving from West to East, is also shown.
The cross marks the Galactic Center.
}
\label{fig_location}
\end{figure*}

\begin{figure}
\includegraphics[width=8.8cm]{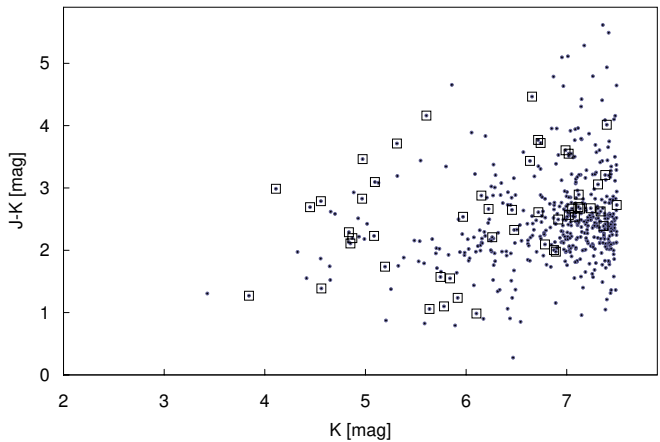}
\caption {Color-magnitude diagram for the sources 
of which we recorded LO events as 
listed in Table~\ref{lo_complete} (open squares).
The small dots mark all sources occulted by the Moon
in the same time span, to the limit K$\le 7.5$\,mag.
}
\label{fig_km}
\end{figure}

We observed a passage of the Moon in a crowded region of the
Galactic Bulge, in the night of March 21-22, 2006. The center
of this region was located at $\approx 17^{\rm h} 42^{\rm m}$
and $-28\degr29\arcmin$. Fig.~\ref{fig_location} shows the
area, with the apparent lunar path superimposed.
It can be seen that, at closest approach, the limb of the Moon
as seen from Paranal reached only a few arcseconds from the
true Galactic Center.
The region is 
heavily reddened by interstellar dust. A near-IR color-magnitude
diagram is shown in Fig.~\ref{fig_km}. Due to extinction,
few sources in this region of the sky have optical counterparts. 
Our LO predictions were based mostly on the 2MASS Catalogue
(Cutri et al \cite{2003yCat.2246....0C}). To the limit K$\le7.5$\,mag
a total of 509 sources were predicted to be 
occulted by the Moon in 5 hours.
Our observations began at 6:11~UT and ended at 10:25~UT. A total
of 51 events were recorded. The efficiency was limited in the first
part of the observations by 
several missed or wrongly recorded events.
In fact, this run represented 
a commissioning of the burst mode at ISAAC, which at the time
had not yet been extensively tested. In the last part of the
observations, with the technical aspects under control, the typical 
interval between recorded events was as short as three minutes and
limited by the telescope pointing and data storing times.

\begin{table*}
\caption{List of the recorded occultation events\label{lo_complete}}
\begin{tabular}{lccrrrlll}
\hline 
\hline 
\multicolumn{1}{c}{2MASS id}&
Frame&
NDIT&
\multicolumn{1}{c}{Time}&
\multicolumn{1}{c}{K}&
\multicolumn{1}{c}{J-K}&
\multicolumn{1}{c}{Sp}&
\multicolumn{1}{c}{Cross-Id} &Notes\\
\multicolumn{1}{c}{}&
(px)&
&
\multicolumn{1}{c}{(UT)}&
\multicolumn{2}{c}{(mag)}&
\multicolumn{1}{c}{}&
\multicolumn{1}{c}{}&
\multicolumn{1}{c}{}\\
\hline
17383314-2836158		&	32	&	14000	&	06:11:57	&	5.64	&	1.06	&	\multicolumn{1}{c}{$-$}	&	\multicolumn{1}{c}{$-$}	&	df, 50\% off edge	\\
17383141-2812445		&	32	&	14000	&	06:15:19	&	6.78	&	2.09	&	\multicolumn{1}{c}{$-$}	&	\multicolumn{1}{c}{$-$}	&	df	\\
17394834-2809136		&	64	&	12000	&	06:23:51	&	6.48	&	2.33	&	\multicolumn{1}{c}{$-$}	&	\multicolumn{1}{c}{$-$}	&	near terminator	\\
17385533-2836105		&	64	&	12000	&	06:28:10	&	7.34	&	2.62	&	\multicolumn{1}{c}{$-$}	&	\multicolumn{1}{c}{$-$}	&	df, 50\% off edge	\\
17385407-2833227		&	64	&	12000	&	06:32:56	&	7.13	&	2.67	&	\multicolumn{1}{c}{$-$}	&	\multicolumn{1}{c}{$-$}	&	df, 50\% off edge	\\
17391567-2826194		&	64	&	12000	&	06:47:58	&	6.46	&	2.65	&	\multicolumn{1}{c}{$-$}	&	\multicolumn{1}{c}{$-$}	&	df	\\
17394022-2820124		&	64	&	12000	&	06:55:09	&	7.12	&	2.69	&	\multicolumn{1}{c}{$-$}	&	\multicolumn{1}{c}{$-$}	&	df, close to edge	\\
17394602-2822089		&	64	&	12000	&	07:00:05	&	7.05	&	2.67	&	\multicolumn{1}{c}{$-$}	&	\multicolumn{1}{c}{$-$}	&	df	\\
17395038-2824061		&	64	&	12000	&	07:04:00	&	5.78	&	1.10	&	\multicolumn{1}{c}{$-$}	&	CD-28 13409	&	 not seen	\\
17410318-2814151		&	64	&	12000	&	07:08:12	&	7.02	&	2.57	&	\multicolumn{1}{c}{$-$}	&	\multicolumn{1}{c}{$-$}	&	edge (75\% off)	\\
17402920-2842361		&	64	&	12000	&	07:14:29	&	7.50	&	2.73	&	\multicolumn{1}{c}{$-$}	&	\multicolumn{1}{c}{$-$}	&	 not seen	\\
17405112-2817130		&	64	&	12000	&	07:18:35	&	4.87	&	2.20	&	M6.5	&	[RHI84] 10-333	&	df	\\
17412116-2816010		&	64	&	12000	&	07:22:54	&	5.84	&	1.55	&	M5	&	[RHI84] 10-356	&	df, 2 stars	\\
17402750-2833099		&	64	&	12000	&	07:26:35	&	7.24	&	2.67	&	\multicolumn{1}{c}{$-$}	&	\multicolumn{1}{c}{$-$}	&	 not seen	\\
17411415-2818051		&	64	&	12000	&	07:29:49	&	6.87	&	2.00	&	M5	&	[RHI84] 10-352	&	df, 2 stars	\\
17404347-2827185		&	64	&	12000	&	07:33:08	&	6.89	&	1.98	&	\multicolumn{1}{c}{$-$}	&	\multicolumn{1}{c}{$-$}	&	df, 50\% off edge	\\
17410937-2822418		&	64	&	12000	&	07:39:35	&	6.92	&	2.49	&	\multicolumn{1}{c}{$-$}	&	DENIS-P J174109.4-282242	&	df	\\
17412349-2821538		&	64	&	12000	&	07:44:18	&	7.37	&	2.39	&	\multicolumn{1}{c}{$-$}	&	\multicolumn{1}{c}{$-$}	&	df	\\
17411109-2840486		&	64	&	12000	&	07:47:31	&	7.31	&	3.06	&	\multicolumn{1}{c}{$-$}	&	\multicolumn{1}{c}{$-$}	&	 not seen	\\
17423294-2818473		&	64	&	12000	&	07:52:21	&	3.84	&	1.27	&	G8Iab	&	HD 160706	&	near terminator	\\
17414641-2822593		&	64	&	12000	&	07:57:10	&	5.09	&	2.23	&	\multicolumn{1}{c}{$-$}	&	ISOGAL-P J174146.5-282301	&	df	\\
17413435-2829225		&	64	&	12000	&	08:01:38	&	4.85	&	2.11	&	\multicolumn{1}{c}{$-$}	&	DENIS-P J174134.4-282922	&	df, 10\% off edge	\\
17414185-2842287		&	64	&	12000	&	08:05:22	&	6.22	&	2.66	&	\multicolumn{1}{c}{$-$}	&	\multicolumn{1}{c}{$-$}	&	 not seen	\\
17415470-2826596		&	64	&	12000	&	08:09:09	&	7.08	&	2.54	&	\multicolumn{1}{c}{$-$}	&	DENIS-P J174154.7-282659	&	df, 95\% off edge	\\
17425620-2820370		&	64	&	12000	&	08:13:32	&	4.56	&	1.39	&	\multicolumn{1}{c}{$-$}	&	2MASS J17425620-2820370	&	 not seen	\\
17415719-2834236		&	64	&	12000	&	08:17:55	&	7.12	&	2.89	&	\multicolumn{1}{c}{$-$}	&	\multicolumn{1}{c}{$-$}	&	very faint	\\
17420509-2833465		&	64	&	12000	&	08:22:10	&	6.15	&	2.88	&	\multicolumn{1}{c}{$-$}	&	\multicolumn{1}{c}{$-$}	&	df, 50\% off edge	\\
17423746-2825311		&	64	&	12000	&	08:27:23	&	5.20	&	1.74	&	\multicolumn{1}{c}{$-$}	&	\multicolumn{1}{c}{$-$}	&	df, very faint	\\
17424039-2826255		&	64	&	12000	&	08:31:04	&	4.56	&	2.79	&	\multicolumn{1}{c}{$-$}	&	IRAS 17395-2825	&	 not seen	\\
17432345-2853503		&	64	&	12000	&	08:38:21	&	4.97	&	3.46	&	\multicolumn{1}{c}{$-$}	&	ISOGAL-P J174323.6-285349	&	 not seen	\\
17432585-2823598		&	64	&	12000	&	08:44:05	&	5.92	&	1.24	&	\multicolumn{1}{c}{$-$}	&	\multicolumn{1}{c}{$-$}	&	df, 15\% off edge	\\
17430791-2828016		&	64	&	12000	&	08:48:53	&	7.02	&	3.55	&	\multicolumn{1}{c}{$-$}	&	\multicolumn{1}{c}{$-$}	&	df	\\
17425497-2836593		&	64	&	12000	&	08:53:18	&	6.72	&	3.77	&	\multicolumn{1}{c}{$-$}	&	\multicolumn{1}{c}{$-$}	&	df, 50\% off edge	\\
17432661-2827590		&	64	&	12000	&	08:58:32	&	6.26	&	2.21	&	M2:	&	[RHI84] 10-443	&	df, 50\% off edge	\\
17430901-2837039		&	64	&	12000	&	09:01:53	&	6.99	&	3.60	&	\multicolumn{1}{c}{$-$}	&	\multicolumn{1}{c}{$-$}	&	df, 30\% off edge	\\
17431348-2837004		&	64	&	12000	&	09:04:34	&	6.74	&	3.72	&	\multicolumn{1}{c}{$-$}	&	\multicolumn{1}{c}{$-$}	&	df	\\
17434830-2828129		&	64	&	12000	&	09:10:32	&	4.84	&	2.29	&	\multicolumn{1}{c}{$-$}	&	\multicolumn{1}{c}{$-$}	&	df, 95\% off edge	\\
17435893-2827521		&	64	&	12000	&	09:15:00	&	5.97	&	2.54	&	\multicolumn{1}{c}{$-$}	&	\multicolumn{1}{c}{$-$}	&	df, second star	\\
17444369-2825027		&	64	&	12000	&	09:17:24	&	7.38	&	3.21	&	\multicolumn{1}{c}{$-$}	&	\multicolumn{1}{c}{$-$}	&	df, 5\% off edge	\\
17435704-2830432		&	64	&	12000	&	09:21:47	&	5.61	&	4.16	&	\multicolumn{1}{c}{$-$}	&	IRAS 17407-2829	&	df, 25\% off edge	\\
17434681-2840176		&	64	&	12000	&	09:26:43	&	4.45	&	2.69	&	M6.5	&	ISOGAL-P J174346.6-284017	&	 not seen	\\
17435399-2841285		&	64	&	12000	&	09:31:28	&	4.11	&	2.99	&	\multicolumn{1}{c}{$-$}	&	IRAS 17407-2840	&	df	\\
17445251-2826479		&	64	&	12000	&	09:36:45	&	5.09	&	3.10	&	\multicolumn{1}{c}{$-$}	&	\multicolumn{1}{c}{$-$}	&	 not seen	\\
17441366-2848501		&	64	&	8000	&	09:39:53	&	6.10	&	0.99	&	K0	&	HD 316221	&	5\% off edge	\\
17443976-2830568		&	64	&	10000	&	09:46:46	&	7.40	&	4.01	&	\multicolumn{1}{c}{$-$}	&	\multicolumn{1}{c}{$-$}	&	 not seen	\\
17452448-2826434		&	64	&	12000	&	09:49:33	&	6.72	&	2.61	&	\multicolumn{1}{c}{$-$}	&	\multicolumn{1}{c}{$-$}	&	 not seen	\\
17450413-2853412		&	64	&	12000	&	10:05:20	&	5.75	&	1.57	&	M2	&	[RHI84] 10-515	&	df	\\
17452968-2829325		&	64	&	12000	&	10:10:43	&	6.64	&	3.43	&	\multicolumn{1}{c}{$-$}	&	\multicolumn{1}{c}{$-$}	&	gf, 2 more stars	\\
17445945-2843238		&	64	&	12000	&	10:14:15	&	4.97	&	2.83	&	\multicolumn{1}{c}{$-$}	&	\multicolumn{1}{c}{$-$}	&	gf	\\
17460150-2828035		&	64	&	12000	&	10:19:24	&	6.66	&	4.47	&	\multicolumn{1}{c}{$-$}	&	\multicolumn{1}{c}{$-$}	&	gf	\\
17453224-2833429		&	64	&	12000	&	10:25:26	&	5.31	&	3.71	&	\multicolumn{1}{c}{$-$}	&	ISOGAL-P J174532.3-283338	&	gf, 5\% off edge	\\
\hline 
\hline 
\end{tabular}
\\
{\scriptsize
Frame is the size of the subwindow and NDIT the number of
frames used in the raw burst mode. The number
of frames in the reformatted FITS cube is NDIT/2-1, and this also
determines the total time span of the recorded data.
In the notes, df stands for defocused and gf for good focus.
}
\end{table*}

The limit K$\le7.5$\,mag was quite arbitrary. In fact, in the same time span
the number of occulted  sources
with K$\la 11$\,mag
was more than 15,000.
As we will show later this magnitude limit would have been realistic,
but it was clearly impossible to observe
hundreds, not to mention thousands, of sources.
Even a manual selection of the best ones would have been a
daunting task, and for this we developed
a prioritization rule which takes into account the magnitude, 
the color (redder objects were given more weight), 
and the time intervals between the events.
We also flagged and gave additional priority to sources with previous
observations or classification avalailable in the SIMBAD database.

The lunar phase was -54\%. In general LO observations are possible
only on the dark lunar limb, thus we observed reappearances:
the telescope was pointed at the nominal star position while this
was still occulted, and data were recorded starting about thirty
seconds before the nominal reapperance time. In reality, predictions
resulted accurate to few seconds in all cases. 
Thanks to the work of Paranal software engineers,
the event times
were included in the files that control telescope and instrument
operation (so-called Observing Blocks or OBs). 
These latter were then loaded well in advance, and
data acquisition would start automatically at the preset time.

\begin{figure*}
\resizebox{\hsize}{!}{\includegraphics{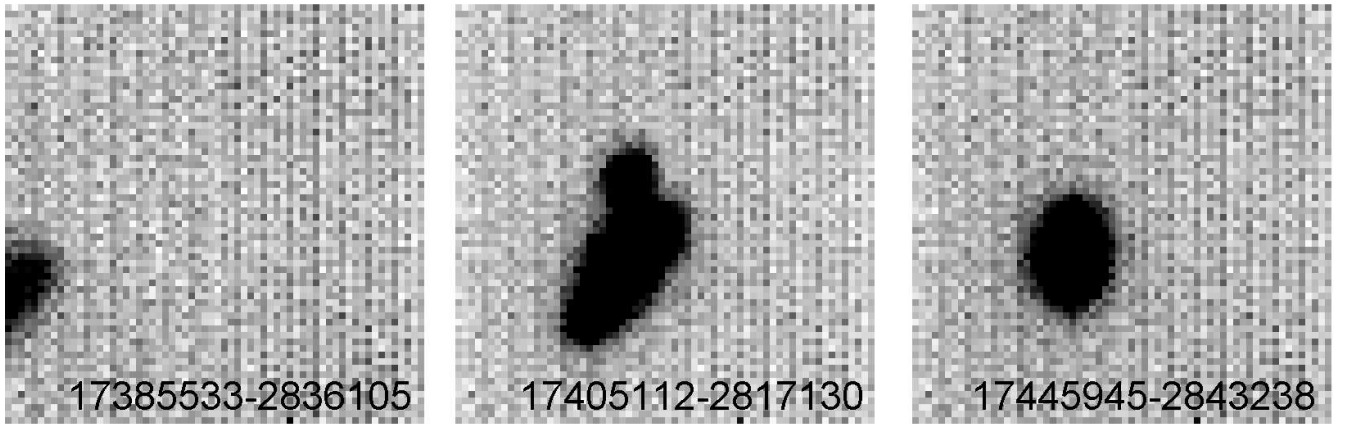}}
\caption {Illustration of the adverse effects of image clipping
due to inaccurate blind pointing (left) and of poor image quality
due to the active optics being switched off (middle, note that
the target is a single star), as described in the text. The right
figure is an example of a more regular case, for comparison.
All three figures are time averages over several thousand frames
on the subwindow of the Aladdin array in ISAAC. The windows are 
$\approx9.5\arcsec$ on the side.
The 2MASS names of the occulted stars are indicated.
}
\label{fig_averages}
\end{figure*}

Two  factors affected significantly the quality of the
resulting data. Firstly, the
active optics corrections of the primary mirror had to be switched off.
As a result, the image quality was rather poor especially in the first
part of the run. Towards the end of the run,
we activated the active optics a few times
in-between events, thus obtaining much better image quality.
Secondly,  
in a few cases part of the signal fell outside the sub-window
(clipping).
Blind pointing was done by means of offsets from a nearby star
but, given also the bad image
quality and the small detector sub-window, it was sometimes not
sufficiently accurate. The sub-window
was kept to 64$\times$64 pixels, or $\approx9.5\arcsec$ squared,
in order to achieve sufficiently fast read-outs.
We stress that in general this does not have too many negative effects
on the occultation light curve except of course for the signal-to-noise 
ratio (SNR). Fig.~\ref{fig_averages} illustrates the
above mentioned effects.

The critical parameter for the angular resolution achieved in a
LO, along with the signal-to-noise ratio (SNR), is the sampling
of the light curve. For this, the burst mode implemented on the
Aladdin detector of the ISAAC instrument was the critical factor,
allowing us to obtain times of just 3.2\,ms on a 32$\times$32 subarray
and 6.4\,ms sampling on a 64$\times$64 subarray.  In order to increase
the accuracy of source centering during blind pointing of the
occulted sources, all the events but the first two were 
recorded in 64$\times$64 mode.
The length of the data stream  was kept between 8000 and 14000 frames
(see Table~\ref{lo_complete}).
We do not describe here the details of the burst mode for
ISAAC, 
for which we refer to Richichi et al. (\cite{vlt_elt}).
The burst mode is also
possible at 
several other ESO instruments both
at Paranal and La Silla (see for example
Domiciano de Souza et al. \cite{burstvisir},
Poncelet et al. \cite{burstagn},
Sicardy et al. \cite{burstnaco}).
The raw data need to be reformatted in
proper FITS files, which are then processed using the AWLORP pipeline.
Schematically, this latter performs both an automated light curve 
extraction based on a
mask estimation which preserves the object-pixel connectivity,
and computes an 
automated estimation of the basic parameters of each light curve,
namely time of occultation, stellar signal,
background signal, and rate of lunar motion. This is achieved via
a novel algorithm based on wavelet-decomposition.
Details on AWLORP can
be found in Fors (\cite{thesis}) and Fors et al. (\cite{fors08}).

Once the data are prepared with the proper structure, AWLORP runs
automatically, saving the user from a lengthy and error-prone
screening and producing preliminary fits and plots
for all light curves.   A total of 30 events were found to have
data of sufficient quality: the remaining 21 were discarded  because
of problems in the pointing or in the data acquisition.
For the data analysis, we employed
the model-dependent fitting ALOR
(see Richichi et al. \cite{richichi96}) and
the model-independent estimation
of the brightness profile CAL (Richichi \cite{CAL}).



\section{Results}\label{results}
The complete list of recorded occultation events is given in
Table~\ref{lo_complete}.
It includes  comments about the quality of the data
and missed events. 
The K-band magnitudes range from 4.1 to
7.4.
In the table, and throughout the paper, we use 2MASS names.
Most of the sources do not have a counterpart in the 
Simbad database. Only 14 of the sources reliably recorded
can be cross-identified with optical or IR sources, and these
are also listed in the table.

\begin{figure}
\includegraphics[width=8.8cm]{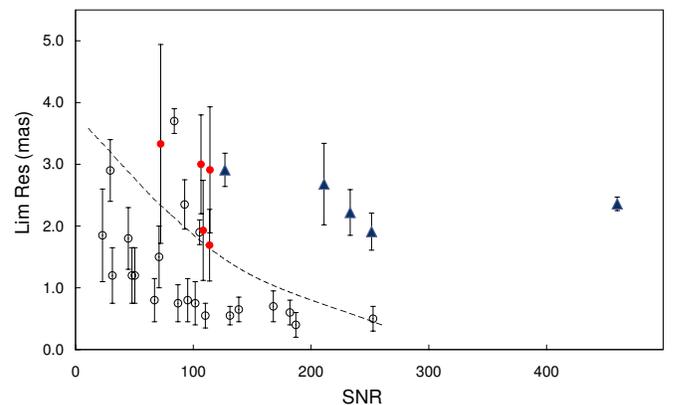}
\caption {Sources for which an upper limit on the angular diameter
could be established using the method described in the text are
marked with outlined circles. The filled circles are sources for
which an angular diameter could be formally established, but for
which the associated errorbar is such that they are undistinguishable
from being unresolved.  The dashed line is an arbitrary representation
of a tentative SNR-limiting angular resolution relationship.
The triangles are sources for which the angular diameter
and associated error lies tentatively or convincingly above
the limiting angular resolution.
}
\label{fig_diam}
\end{figure}

We have tested the resolved/unresolved character of all the
sources, using the method first outlined  by Richichi et al.
(\cite{richichi96}) and applied in some of our previous papers
such as in Fors et al. (\cite{fors04}) and Richichi et al.
(\cite{ca2006}). The results are shown in Fig.~\ref{fig_diam},
and we discuss below separately the resolved
and the unresolved sources.

We show in the following just two 
examples of recorded light curves along with their best 
fits.  The complete set of light curves and fits is available online.
In spite of the previously mentioned problems, we note that  our
sample includes some of the best LO data ever recorded, both in terms
of sensitivity and of SNR.  The
performance of the 8.2\,m VLT telescope for LO work proved to be
outstanding,
not just due to its light collecting power but also to the great
reduction of atmospheric scintillation - which is often the main
limiting SNR factor for LO. In terms of angular resolution, the
relationship shown in Fig.~\ref{fig_diam} is consistent
with that established at other, smaller telescopes for which
the detector integration time (DIT)
was considerably shorter (see Richichi et al. \cite{richichi96},
\cite{ca2006}).
We note that the SNR of the fit shown in Fig.~\ref{fig_gc477}
is, as far as we know, the best ever achieved for a LO light curve.
Just a few milliarcseconds away from the main star
it would have been possible to
detect a companion fainter than the primary by
7.8 and 7.1\,mag in
the occulted and unocculted part respectively, i.e. on 
opposite sides of the star.
This compares favorably with the best AO-assisted high-contrast
imaging close to a bright star on this range of separations
(Masciadri et al. \cite{masciadri05}).

\subsection{Resolved sources}
We have found two sub-arcsecond (projected) separation binaries,
two stars with a convincingly resolved angular size, and
three other stars which we consider as marginally resolved.
These sources are listed in Table~\ref{tab:results},
using the same format already used in Richichi et al.~(\cite{ca2006})
and other papers in that series.
In summary, the columns list the absolute value of the fitted linear rate of
the event V, its deviation from the predicted rate V$_{\rm{t}}$,
the local lunar limb slope $\psi$, the position and
contact angles, the signal--to--noise ratio (SNR). For binary detections, the
projected separation and the brightness ratio are given, while for resolved
stars
the angular diameter $\phi_{\rm UD}$ is reported, under the assumption of
a uniform stellar disc. 
All angular quantities are computed from the fitted
rate of the event.
\begin{table*}
\caption{Summary of results.}
\label{tab:results}
\centering          
\begin{tabular}{lcrrrrrrrr}
\hline 
\hline 
\multicolumn{1}{c}{(1)}&
\multicolumn{1}{c}{(2)}&
\multicolumn{1}{c}{(3)}&
\multicolumn{1}{c}{(4)}&
\multicolumn{1}{c}{(5)}&
\multicolumn{1}{c}{(6)}&
\multicolumn{1}{c}{(7)}&
\multicolumn{1}{c}{(8)}&
\multicolumn{1}{c}{(9)}&
\multicolumn{1}{c}{(10)}\\
\multicolumn{1}{c}{Source}&
\multicolumn{1}{c}{$|$V$|$ (m/ms)}&
\multicolumn{1}{c}{V/V$_{\rm{t}}$--1}&
\multicolumn{1}{c}{$\psi $($\degr$)}&
\multicolumn{1}{c}{PA($\degr$)}&
\multicolumn{1}{c}{CA($\degr$)}&
\multicolumn{1}{c}{SNR}&
\multicolumn{1}{c}{Sep. (mas)}&
\multicolumn{1}{c}{Br. Ratio}&
\multicolumn{1}{c}{$\phi_{\rm UD}$ (mas)}\\
\hline 
17391567-2826194 & 0.7702 & $-$2.9\% & $-$7 & 268 & 163 & 211.0 &    &    & 2.68 $\pm$ 0.66 \\
17412116-2816010 & 0.3245 & 5.7\% & $-$2 & 350 & 245 & 105.4 & 711.7 $\pm$ 3.4 & 22.6 $\pm$ 0.8 &    \\
17435893-2827521 & 0.4688 & 4.9\% & $-$3 & 321 & 222 & 233.2 &    &    & 2.22 $\pm$ 0.37 \\
17444369-2825027 & 0.1802 & 62.3\% & \multicolumn{3}{c}{see text}     & 68.2 & 171.6 $\pm$ 25.8 & 47.8 $\pm$ 7.5 &    \\
17435399-2841285 & 0.6095 & $-$1.4\% & $-$5 & 266 & 168 & 460.0 &    &    & 2.36 $\pm$ 0.11 \\
17445945-2843238 & 0.5945 & $-$1.7\% & $-$7 & 263 & 168 & 251.4 &    &    & 1.91 $\pm$ 0.30 \\
17453224-2833429 & 0.5328 & 7\% & $-$6 & 302 & 209 & 126.9 &    &    & 2.9 (shell) \\
\hline 
\hline 
\end{tabular}
\end{table*}

Several of our acquisition images show companions,
but we report only two binaries which
have projected separations
below one arcsecond. This is rather large for the context of
LO binary detections. In fact, for separations 
$\ga 0\farcs5$ the LO technique is relatively inaccurate, due to
possible differences in the local limb slope at the points of occultation
of the two components. The two objects
are
{\object 2MASS 17412116-2816010} 
and
{\object 2MASS 17444369-2825027}.
They are most likely projection and not physical
pairs: we report them here for completeness, since
no previous mention was present in the literature.
The former
was classified by Raharto et al.
(\cite{rhi84}) as a M5 star. 
We note that the difference between predicted and observed
lunar rate for the second star
was quite significant, as reported
in Table~\ref{tab:results}. The contact angle for this event
was only $260\degr$, i.e. nearly grazing, and the predicted
lunar rate only $-0\farcs1$/s. Under these conditions, it is not
uncommon that either a small error in the predictions or a local
limb slope can significantly  alter the computation of the 
effective PA and CA.

\begin{figure}
\includegraphics[width=8.8cm]{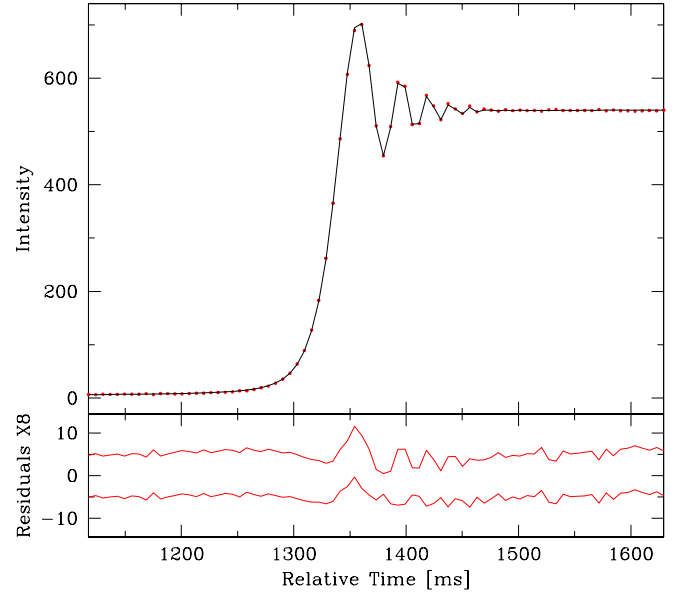}
\caption {{\it Top:} data (dots) and best fit (solid line)
for
{\object 2MASS 17435399-2841285}.
{\it Bottom:} 
enlarged by eight and displaced by arbitrary
offsets for clarity, the residuals of the fits
with a point-like source (above, $\chi^2=2.0$) and with
a uniform-disk model, resulting in a diameter of $2.36\pm0.11$\,mas
(below, $\chi^2=1.1$).
}
\label{fig_gc477}
\end{figure}

The remaining stars in Table~\ref{tab:results} are those which
appear significantly different from point-like. 
The two most interesting ones are 
{\object 2MASS 17435399-2841285} 
and
{\object 2MASS 17453224-2833429}, which happen to be
both masing AGB stars.
Unfortunately the distance determination for these stars
is uncertain, especially in the absence of detailed photometric
and pulsational properties.
The former  
has an angular size which, albeit small in absolute terms,
appears to be convincingly established (see Fig.~\ref{fig_diam}
and Fig.~\ref{fig_gc477}).
This star is cross-identified with {\object IRAS 17407-2840},  
a SiO maser previously measured by
Messineo et al. (\cite{messineo02},
\cite{messineo04},
\cite{messineo05}).
Its radial velocity  is negative and cannot be used for a kinematic
distance, but Messineo (priv. comm.) estimates a distance 
of 3.7\,kpc  based on the average extinction and the model
of Drimmel (\cite{drimmel}). According to this,
the angular size listed in Table~\ref{tab:results} would translate
to over 8\,AU. Apart from consideration on the distance uncertainty,
it is likely that the LO-derived size is not indicative of an
uniform-disk diameter but rather a mix of the photosphere and
the immediate circumstellar emission.

\begin{figure}
\includegraphics[width=8.8cm]{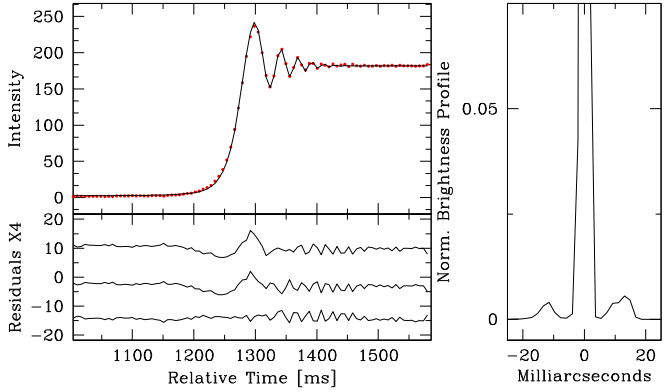}
\caption {{\it Left:} upper panel, data (dots) and best fit (solid line)
for
{\object 2MASS 17453224-2833429}.  The lower panel shows, on a scale
enlarged by four  and
displaced by arbitrary offsets
for clarity, the residuals of three different fits.
Top, fit with a point-like source ($\chi^2=7.0$). Middle, fit with
a uniform-disk model, resulting in a diameter of $2.91\pm0.27$\,mas
($\chi^2=6.3$). Bottom, fit with the model-independent CAL method
($\chi^2=1.6$), by far the most convincing.
{\it Right:} brightness profile reconstructed by the CAL method,
corresponding to the residuals of the lowermost fit on the left.
}
\label{fig_gc509}
\end{figure}

{\object 2MASS 17453224-2833429}
has a similar nature and set of references, with the
addition  of
Sjouwerman et al. (\cite{Sjou04}). The fit for this source
gave a SNR of 127 which is lower than expected for the
magnitude of this source (K=5.1\,mag), and the UD size
reported in Table~\ref{tab:results} should be considered
only as an indication.  A model-independent analysis by CAL
revealed a faint but well defined 
extended emission around the central component,
as shown in Fig.~\ref{fig_gc509}. The profile of this emission
is almost symmetrical, has the rim brightening
typical of optically thin shells
and is consistent with the masing nature of this star.
In this case both a kinematic  and an extinction distance
can be estimated, at 4.5 and 3.3\,kpc respectively
(Messineo, priv. comm.). By using their average,
the inner shell radius of 10-15\,mas would translate to $\approx$40\,AU.
We note that the residuals of the uniform-disk model for
{\object 2MASS 17435399-2841285} indicate the possibility 
of a circumstellar shell being detected also in this case
(cfr. Fig.~\ref{fig_gc477}), but given that the $\chi^2$ was
already significantly good we did not proceed with more
extreme fits.
The different 
signatures of the shell for these two maser stars 
could be related to their difference in
J-K color (3.0 versus 3.7\,mag).

The remaining three stars in 
Table~\ref{tab:results}
appear resolved, although their location above the 
SNR-resolution relationship of Fig.~\ref{fig_diam}  is less
definite than for the two maser stars just mentioned and we
prefer to consider their diameters as tentative for the moment.
There are no known cross-identifications or literature entries
for these stars.

\subsection{Unresolved stars and performance}
Most of the stars in our sample appear unresolved, as is to be
expected from random observations of field stars.
Richichi et al. (\cite{ca2006}) have discussed the incidence
of binary stars in such samples for LO observations based on
the 2MASS catalogue, and our present results are consistent
within the uncertainty of small statistics.
The unresolved stars are also useful, namely in that the
relationship of Fig.~\ref{fig_diam} establishes rather strict
upper limits on their size and on the brightness ratio of 
unseen components, if any. Thus these stars can be safely
used as calibrators for long-baseline interferometry.
\begin{figure}
\includegraphics[width=8.8cm]{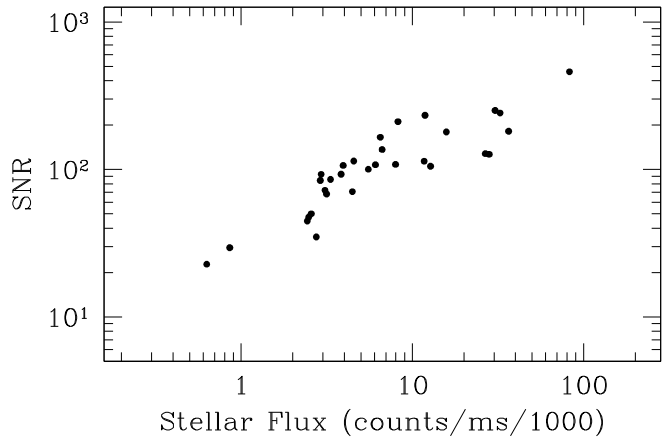}
\caption {
Plot of the SNR achieved in our fits of both resolved and
unresolved source, as a function of the total measured flux.
}
\label{fig_f0snr}
\end{figure}

\begin{figure}
\includegraphics[width=8.8cm]{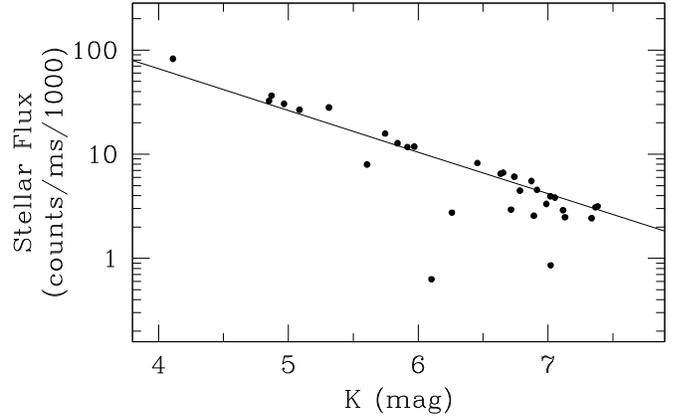}
\caption {
Plot of the total measured flux against the K magnitude of the
sources (from 2MASS). 
The line shows the expected counts, based on the ISAAC
Exposure Time Calculator.
Note the discrepant cases, due to the problems mentioned in the
text.
}
\label{fig_kf0}
\end{figure}

Further, the unresolved stars allow us to
investigate the performance of the burst mode at the VLT for LO,
similar to what was done for other instrument and telescope
configurations in the past (see Fors et al. \cite{fors04}).
In Fig.~\ref{fig_f0snr} we plot the SNR achieved in the fits of
all sources (the unresolved ones being the majority), against
the total measured flux. This is a better quantity to use than
the actual magnitude, since as mentioned earlier many of our sources
were recorded at the edge of the subarray field with considerable
loss of signal. Also intrinsic variability might be a cause
for some of the stars. Indeed, Fig.~\ref{fig_kf0}
shows many outliers from the ideal K magnitude-observed flux
are present. Using the measures signal instead, 
Fig.~\ref{fig_f0snr} shows a rather linear relationship across 
over two orders of magnitudes in flux. This implies that at
the faint end we were still in a photon-limited regime, and at
at the bright end the incidence of scintillation and possible
non-linearities was not yet significant. We estimate that non-linearity of
the detector should become important above K$\approx$3\,mag, i.e. one
magnitude above our brightest target for this run.

In spite of the outlyers, the relationship between K magnitude
and counts clearly emerges from 
Fig.~\ref{fig_kf0}, and it is in excellent agreement with the
expected ISAAC performance.
Comparing the two figures, we estimate at first approximation that
under the same conditions of sampling and DIT (6.4\,ms) it should
have been possible to obtain SNR=1 for K$\approx$12\,mag. 
This is the
minimum SNR necessary to detect a faint companion in a LO light curve,
although measurements of angular diameters or extended emission would
require SNR$\ga$10.
We note that, at the faint magnitude end, critical parameters are the
diffuse scattered light background, 
which is a function of the lunar phase and distance of the
occultaton point from the terminator, and the pixel scale. 
Ideally, this latter
should match the seeing disc, but in our case with a pixel scale
of about $0\farcs148$ we significantly oversampled the background.

\onlfig{9}{
\begin{figure*}
\resizebox{\hsize}{!}{\includegraphics{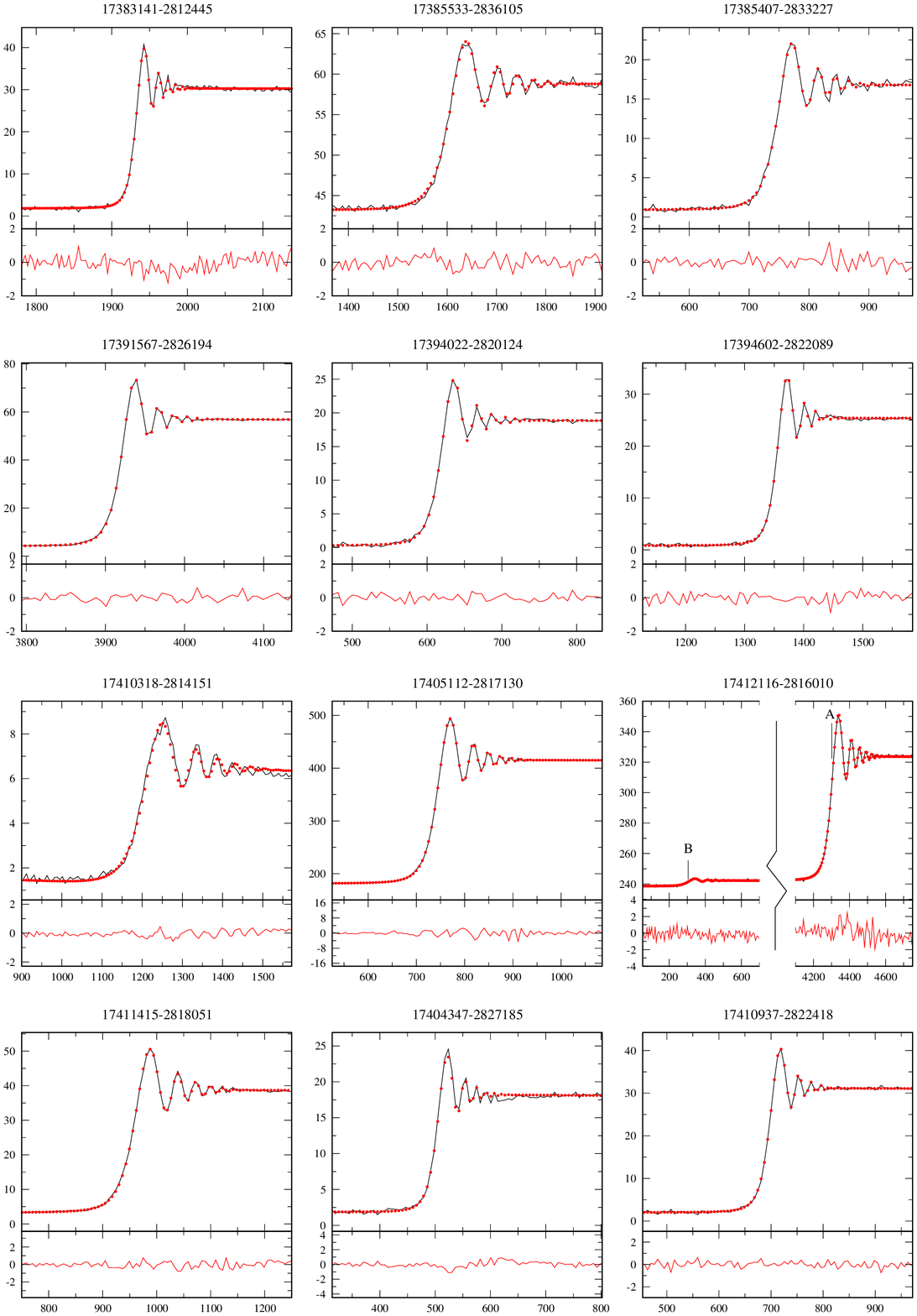}}
\caption {Light curves of observed LOs}
\label{fig:LO}
\end{figure*}
}

\onlfig{9}{
\begin{figure*}
\resizebox{\hsize}{!}{\includegraphics{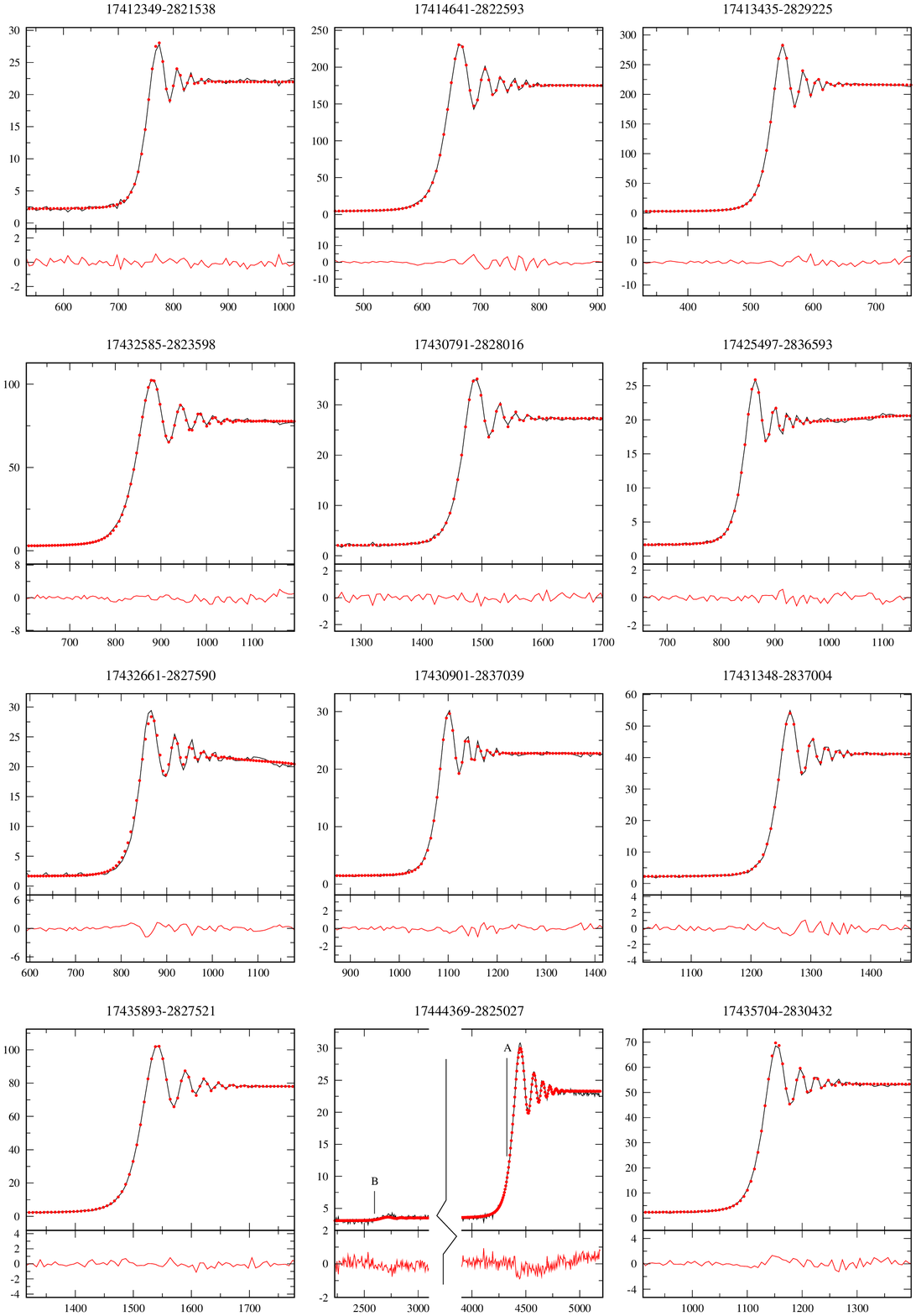}}
\caption {Light curves of observed LOs (continued)}
\end{figure*}
}

\onlfig{9}{
\begin{figure*}
\resizebox{\hsize}{!}{\includegraphics{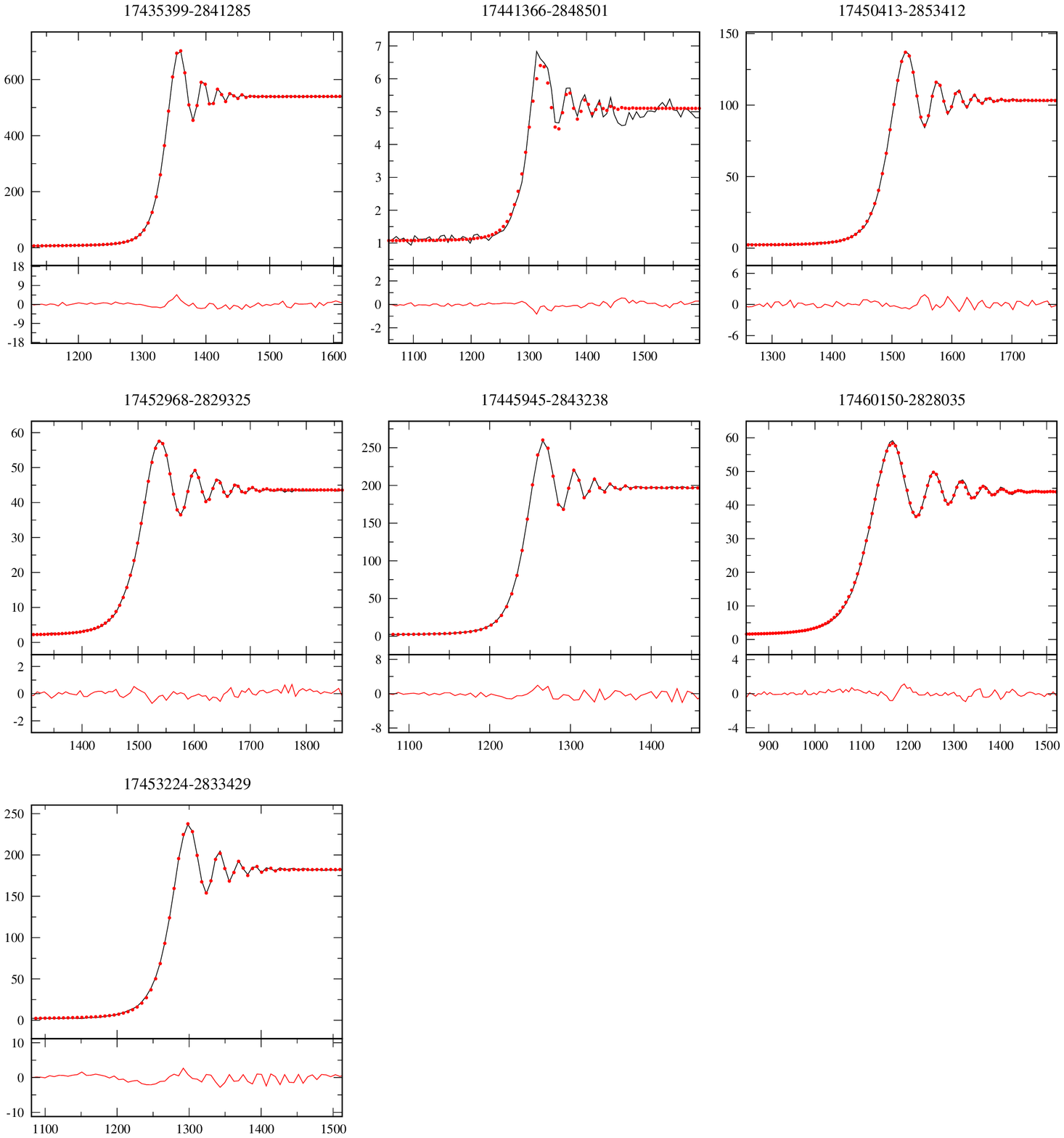}}
\caption {Light curves of observed LOs (continued)}
\end{figure*}
}


\section{Conclusions}
We have presented the first LO results obtained using the so-called
burst mode of the ISAAC instrument at the ESO Antu VLT telescope.
This mode permits to record data streams  on a small subarray
at high temporal frequency
(3.2\,ms, although mostly 6.4\,ms were used for the present work).
We recorded 51 occultation events over about four hours during
a close approach of the Moon to the Galactic Center. The events
were reappearances from behind the dark limb. This, coupled with 
initial problems of focussing and image quality due mostly to
the commissioning nature of the run, resulted in several sources
being missed or without sufficient quality. 
Nevertheless, 30 of the events resulted in light curves of good,
or even excellent quality, including some of the best SNR traces
ever recorded with this technique. We have revealed two binaries,
three stars with a marginally resolved angular diameter of the order
of 2\,mas, and two resolved masing AGB stars which appear to be
in the foreground of the galactic bulge.
For one of these,
{\object 2MASS 17453224-2833429}, we were able to recover the
brightness profile of the extended emission. This has the typical
signature of an optically thin shell, with inner radius of order
10-15\,mas, corresponding to about 40\,AU.
Follow-up studies are required to obtain further information on
these stars, given that they are all heavily reddened and most have
no known counterpart or literature entries.

Our analysis, including the unresolved sources, has established
an unprecedented
performance of the LO technique in the near-IR using 
a very large telescope and a fast read-out mode.
We find that the angular resolution varies between 4 and 1\,mas, and
even less in the case of very high SNR. Data quality is generally
impressive, compared to smaller telescopes, mostly due to the 
scintillation reduction effect of the large mirror.
The dynamic range is also significantly improved with respect to previous
observations, and we have shown
that in some cases it would have been possible to detect
a companion fainter than the primary by 
about 8 magnitudes in broad K band on angular scales
comparable to the Airy disk of the telescope. This result compares
favorably with other AO-assisted high contrast imaging techniques.

The random selection of sources to be occulted and the fixed-time
nature of the events remain the main downsides of the LO technique.
However, thanks to the very short telescope time required for each
observation, LO are an ideal filler program for times when the telescope
is idle or other programs cannot be conveniently executed.
The service mode observations implemented at the VLT provide the
right context for this, and we have obtained approval for such a
filler program during ESO observing periods 80 and 81 (from October 2007
through September 2008). Several hundreds of OBs have been prepared
for each period, and are in the queue awaiting execution whenever five
minutes of telescope time become available.


\begin{acknowledgements}
This work is partially supported by the 
\emph{ESO Director General's Discretionary Fund}
and by the
\emph{MCYT-SEPCYT Plan Nacional I+D+I AYA \#2005-08604}.
AR wishes to thank the National Astronomical Observatory of
Japan in Mitaka, and Dr. J.~Nishikawa in particular, for their
hospitality during a prolonged stay which led to the final
writing of this paper.
We thank Dr. M.~Messineo for private communications on
the properties of the SiO masers reported in this paper.
This research has made use of the SIMBAD database,
operated at CDS, Strasbourg, France.
\end{acknowledgements}

\end{document}